\documentclass[10pt,dvipsnames]{article}

\usepackage{amsmath,amssymb,here,verbatimbox}
\usepackage[utf8]{inputenc }
\usepackage{amsmath,amssymb,xcolor,bm}
\usepackage{hyperref}
\usepackage{tikz}
\usetikzlibrary{arrows,shapes,calc}
\usepackage{pgfplots}
\usetikzlibrary{positioning} 

\newcommand{\pac}{{\sl pdf-mpc} }
\newcommand{\matlab}{{\sc Matlab} }
\newcommand{\userode}{{\sl \color{BrickRed} user\_ode} }
\newcommand{\userocp}{{\sl \color{BrickRed} user\_ocp} }
\newcommand{\cp}{{\sl \color{Black} create\_solution} }
\newcommand{\useruparam}{{\sl \color{BrickRed} user\_control\_profile} }
\newcommand{\pode}{{\sl \color{Blue} p\_ode} }
\newcommand{\pocp}{{\sl \color{Blue} p\_ocp} }
\newcommand{\puparam}{{\sl \color{Blue} p\_uparam} }
\newcommand{\bmu}{{\bm u}}

\title{{\bf The {\sc \Huge pdf-mpc} package}: \vskip 2mm
\large A Free-Matlab-Coder package for Real-Time Nonlinear Model Predictive Control}
\author{Mazen Alamir \\ \small CNRS University of Grenoble-Alpes}

\begin{document}
\maketitle
\hrule
\begin{abstract}
\noindent This paper describes the Parametrized Derivative-Free Model Predictive Control ({\sl pdf-mpc}) package, a \matlab coder-based set of subroutines that enables a model predictive control problem to be defined and solved. the \pac is made available for free download and use through the website of the author\footnote{http://www.mazenalamir.fr (software section)}.
\end{abstract}
\tableofcontents
\section{Introduction}
Model predictive control \cite{Mayne2000} is probably the most attractive control design methodology nowadays. This is due to its ability to handle constraints, nonlinearity and performance/cost trade-offs. This paper assumes that the reader is familiar with MPC as a control design methodology although a brief recall is provided for the sake of notation. The paper focuses on the way the \pac package can be used to solve the associated problem and to provide the MPC-feedback law in a usable form. \ \\ \ \\ 
As far as {\bf Nonlinear} MPC design is concerned, there are some other packages that are freely available such as {\sc acado} \cite{Houska2011a} (see also the {\sc acado} project website at http://acado.github.io), {\sc mpc-tools} (see the project website at http://jbrwww.che.wisc.edu/software/mpctools/index.html), Chalmers university NMPC software (http://publications.lib.chalmers.se/records/fulltext/146434.pdf) to cite but few ones. \ \\ \ \\
The main reason for which I decided to make the \pac package available for free is that I prefer to use it myself despite of all other existing packages for some reasons I will shortly explain, so maybe some other researchers would share these reasons  and prefer to use the \pac package. More precisely:
\subsection{Main features}
\begin{itemize}
\item[$\checkmark$] \pac package is simply a sort of {\sc Matlab} toolbox ! you do not need to learn new conventions in building models. You are not limited in the number of .m files you need to build your description. All \matlab subroutines that are compatible with {\sc Matlab-coder} are available for to be used in the construction of the problem's components. By using the {\sc matlab-coder} to build a mex-file containing the NMPC feedback solution, the \pac packages combine the natural use of Matlab with the high performance of a compiled solution. The resulting computation times are quite comparable to the best available ones with the flexibility on the top of it ! 
\item[$\checkmark$] \pac package offers a complete freedom in the definition of the decision variables. This feature is well known within the MPC community as the {\bf control parametrization} \cite{alamir:hal-00113043}. Most (if not all) of existing packages exclusively use the standard piece-wise constant control parametrization, ending by a number of decision variables equal to $T/\tau$ where $T$ is the prediction horizon while $\tau$ is the control sampling period. For real-time applications, this is not necessarily the best choice. Moreover, for many problems, the decision variable vector might contain heterogenous components including the instants where the control becomes constant, the prediction horizon itself, the pulsation modes of the Fourier series over which the control profile is defined, the decrease rate of the exponential basis if any, to cite but few examples. The use cases proposed in this paper enable this feature to be clearly understood\footnote{A more extensive set of examples can be examined in \cite{alamir2013pragmatic}.}. 
\item[$\checkmark$] The \pac package does not restrict the definition of the cost function to be of a specific structure. In particular, the cost function is not necessary constrained to be the sum of point-wise defined stage cost terms plus a terminal penalty. \pac  package offers the possibility to define any function of the state and control {\bf trajectories}. This can include a term that penalizes the maximum value of some function of the state/control, a penalty on the state excursion with dead-zone. $\ell_i$-norms can be used with $i\in \mathbb N$. The same comment holds for the definition of the constraints. Note however that the \pac package does not guarantee the convergence of the solver to {\em the} solution (if any) to the resulting optimization problem (and few solvers can pretend doing it for realistic set of initial guesses). However, there is no structural restrictions that prevent such formulations from being used, tested and tuned. 
\item[$\checkmark$] The \pac package offers a simple and intuitive way to distinguish between the {\em real}-life model and the {\em nominal} model used to construct the MPC feedback. This makes easy to check the robustness of the feedback to parameter uncertainties. Moreover, it offers the possibility to use lower order integration schemes (faster computation) in the control design and test the resulting controller on a highly precise simulation model to tune the computational burden. 
\item[$\checkmark$] The \pac package offer a real-time interruptible MPC in the sense that the user defines in advance the maximum number of iteration\footnote{Throughout the paper, a single iteration is a single evaluation of the cost/constraints pair. This is also called a single iteration.}. Moreover, it offers an a priori tight estimation of the cost associated to a single iteration for the specific machine on which it runs. Combining these two information, the user can finely tune the maximum number of iterations given the targeted control updating frequency. This is a valuable action towards the success and stability of real-time MPC \cite{alamir2016}.
\end{itemize} 
To summarize, I believe that for any specific example, one of the existing softwares would perform better that the \pac package but it remains that what \pac can do is not entirely included in any single existing software. I hope that the present paper and the future use of the software by researcher and practitioners can convince can support this {\em claim}.  
\subsection{Why \matlab?}
The reason for this choice is that there are so many people having real-life research problems for which they would like to try nonlinear MPC and who are uneasy with C++, C\# or that sort of tools and who are uneasy with complex installation and path-parametrization steps. These people would welcome a decompressable zip file containing ready to use {\sc matlab} set of subroutines helping them in easy intuitive definition and solution of their MPC problem with an output subroutine that can be immediately {\em integrated} in their \matlab complete solution to their problems. Beyond this pragmatic consideration, it is a fact that using the {\sc Matlab-coder} toolbox enables the creation of C-libraries, dlls although this possibility is not yet offered in the first release of the \pac package which is restricted to the creation of {\sc mex}-function version of the MPC feedback solution.   
\subsection{Contents of this paper}
This manual is organized as follows:
\begin{itemize}
\item[$\diamondsuit$] First of all, the basics of MPC design are recalled in Section \ref{basicsMPC}. The objective here is more to introduce the components of an MPC design problem than to give a self-contained introduction on MPC. Readers who feel uneasy with that level of presentation can refer to \cite{Mayne2000} for more rigorous introduction to MPC.  This section describes the components of an MPC design problem while introducing in parallel the notation used in the \pac package. 
\item[$\diamondsuit$] Section \ref{secstructure} describes the structure of the \pac package that can easily define an MPC design problem such as the ones described in Section \ref{basicsMPC}.
\item[$\diamondsuit$] Section \ref{somimpo} underlines some important facts to be kept in mind when using the \pac package in order to avoid common errors during the creation of the {\sc mex}-function representing the MPC feedback.   
\item[$\diamondsuit$] Section \ref{secexamples} shows some case studies that illustrate the use of the \pac package. 
\item[$\diamondsuit$] The Appendices \ref{asecdownload} and \ref{asecterms} discusses the download/installation procedure and the terms of use of the \pac package. 
\end{itemize} 
\section{MPC design: Recalls and notation}\label{basicsMPC} 
\subsection{The dynamics}
The \pac package is concerned with dynamic systems that are governed with the following class of Ordinary Differential Equations (ODEs):
\begin{equation}
\dot x=\mbox{\rm \userode}(x,u,\mbox{\rm \pode})  \label{system} 
\end{equation} 
where the following notation is used:
\begin{tabbing}
\hskip 15mm \= \hskip 20mm \kill 
$x$ \> :the state vector ($\in \mathbb{R}^{n_x}$)\\
$u$ \> :the control vector ($\in \mathbb{R}^{n_u}$)\\
\pode \> :the structure containing the parameters involved in the definition of the dynamics\\
\userode \>:the map that defines the r.h.s of the dynamics (the name \userode must be respected)
\end{tabbing}
Given a sampling period $\tau>0$ and a prediction horizon of length $T=N\tau$ for some $N\in \mathbb N$, it is common to denote the state trajectory of (\ref{system}) starting from some initial state $x_0$ and under the piece-wise constant control $\bm u:=(u_0,\dots,u_{N-1})\in \mathbb R^{n_u}\times \dots\times \mathbb R^{n_u}$ by:
\begin{equation}
\bm x^\bmu(x_0)=:(\bm x^\bmu_1(x_0),\dots,\bm x^\bmu_N(x_0))\in \mathbb{R}^{n_x}\times \dots\times \mathbb R^{n_x}
\end{equation} 
As it is explained later, the use of the field p\_ode.w enables to introduce uncertainties in a quite flexible way. The computation of the optimal control uses the nominal model $w=0$.
\subsection{The control parametrization}
As it is mentioned in the introduction, the majority (if not all) of existing formulations consider all the components of $\bmu$ as decision variables. Everybody knows however (and many researchers use it in their applied implementation works) that it is sometimes more appropriate to consider a parametrization of the form:
\begin{equation}
\bmu(p):=\mbox{\rm \useruparam}(p,\mbox{\rm \pode},\mbox{\rm \puparam})  \label{uparam} 
\end{equation}  
where the following notation is used:
\begin{tabbing}
\hskip 18mm \= \hskip 20mm \kill 
$p$ \> :the the vector of degrees of freedom,\\
\pode \> :the structure of parameters described above and involved in the definition (\ref{system}) of the dynamics,\\
\puparam \> :the structure containing the parameters that define the control profile over the prediction horizon.\\
$\bm u$ \> The matrix defining the control profiles over the prediction horizon of length p\_uparam.Np.
\end{tabbing}
Note that the control profile can be dependent on the initial state $x_0$. This is because \pode contains the initial state as a required field\footnote{See the list of required fields of all the structures \pode, \puparam and \pocp in Section \ref{secreqfields}.}. The parametrization can also be time-varying provided that the time is a field of \pode or \puparam. This will be more clear through the case studies provided in the paper. \ \\ \ \\ 
Revisiting the notation of the preceding section and as far as the control profiles are parametrized by $p$ according to (\ref{uparam}), the state trajectories depend now on $x_0$ and $p$ so that the following {\em abuse of notation} can be used:
\begin{equation}
\bm x^p(x_0)=:(\bm x^p_1(x_0),\dots,\bm x^p_N(x_0))\in \mathbb{R}^{n_x}\times \dots\times \mathbb R^{n_x}
\end{equation} 

\subsection{The optimization problem}
The optimization problem is defined through a cost function and a set of constraints to be satisfied. The cost function is defined for a given pair $(\bm x,\bmu)$ of state and control {\bf trajectories}. When these trajectories are given for a specific value of the vector of degrees of freedom $p$, The cost function becomes a function of $p$ and, among other parameters, on the initial state value $x_0$. The same comment holds for the definition of the constraints to be satisfied. 
\\ \ \\  
The \pac package enables to handle cost functions of the form:
\begin{equation}
[J,g]=\mbox{\rm \userocp}(\bm x,\bmu,\mbox{\rm \pode,\puparam,\pocp}) \label{theocp} 
\end{equation} 
where the following notation is used:
\begin{tabbing}
\hskip 18mm \= \hskip 20mm \kill 
$\bm x$ \> :the state trajectory,\\
$\bmu$ \> :the control trajectory,\\
\pode \> :the structure of parameters involved in the definition of the dynamics (see above),\\
\puparam \> :the structure of parameters involved in the definition of the control parametrization (see above),\\
\pocp \> :the structure of parameters involved in the definition of the optimization problem (cost+constraints), \\
$J$ \> :the value of the cost function to be minimized by appropriate choice of $p$ (scalar),\\
$g$ \> :this is a scalar that summarizes the satisfaction of the set of constraints.  
\end{tabbing}
Note that while $g$ is a scalar, as many constraints $c_i(p)\le 0$ as required can be handled by simply defining $g$ using one of the following definitions: 
\begin{equation}
g(p):=\max_i(c_i(p))\quad \mbox{\rm or}\quad g(p):=\sum_i\max\{0,c_i(p)\}^q\quad \mbox{\rm etc.} 
\end{equation} 
\subsection{The MPC feedback}
When using the control parametrization defined by (\ref{uparam}) to generate the trajectories $\bm x^p$ using the control profile $\bmu(p)$, the cost function $J(p)$ and the constraints function $g(p)$ become functions of $p$ thanks to (\ref{theocp}), namely:
\begin{equation}
[J(p),g(p)]=\mbox{\rm \userocp}(\bm x^p,\bmu(p),\mbox{\rm \pode,\puparam,\pocp}) 
\end{equation} 
This enables the following optimization problem to be defined:
\begin{equation}
\min_{p\in [p_{min},p_{max}]} J(p)\quad \vert \quad g(p)\le 0 \label{optimpb}  
\end{equation} 
where 
\begin{tabbing}
\hskip 18mm \= \hskip 20mm \kill 
$p_{min}$ \> :the lower bound on the decision variable $p$ ($\in \mathbb{R}^{n_p}$),\\
$p_{max}$ \> :the upper bound on the decision variable $p$ ($\in \mathbb{R}^{n_p}$)
\end{tabbing}
Let us denote by $p^*$ the optimal solution of the optimization problem (\ref{optimpb}). This optimal solution corresponds to the optimal control trajectories defined through (\ref{uparam}) by:
\begin{equation}
\bmu(p^*):=(\bmu_1(p^*),\dots,\bmu_{N}(p^*))\in \mathbb{R}^{n_u}\times \dots\times \mathbb{R}^{n_u}
\end{equation} 
The MPC feedback is defined by the receding-horizon principle in which the first control vector in the optimal sequence, namely $\bmu_1(p^*)\in \mathbb{R}^{n_u}$ is applied to the system during the next sampling period:
\begin{equation}
\mbox{\rm MPC feedback}:\qquad  u:=\bmu_1(p^*)\in \mathbb{R}^{n_u}
\end{equation} 
at the next decision instant, the parameters of the problem are updated (in particular, the current state takes the status of the initial state), a new problem is formulated and solved to get the new optimal sequence of which only the first control is applied during the next sampling period and so on. \ \\ \ \\ 
Obviously this brief presentation hides many real-time problems. In particular, the time needed to solve the optimization problem has to be taken into account. This is done by anticipating what would be the future {\em initial state} and by starting the computation of the optimal solution while applying the control computed in the previous sampling period. This is known in the literature as the preparation step \cite{diehl2005real,Zavala200986}. On the other hand, we referred to the so-called optimal solution in the description of the MPC principle, unfortunately, the optimal solution can never be rigorously obtained. Instead, a stopping condition should be defined and only a sub-optimal solution is delivered. The choice of this stopping condition is a matter of active research area \cite{alamir2016,ALAMIR201565,Alamir_ECC2013}. These advanced topics are not addressed here. However, it is claimed here (and I hope the examples hereafter can make the reader feel it clearly) that the parametrized framework that can be handled by the \pac package is particularly suitable to handle such issues in an easy and efficient way. \ \\ \ \\ 
The following section describes the modules of the package before an example is given to show how they can be used to solve the kind of MPC design problems sketched above. 
\section{The structure of the \pac package} \label{secstructure} 
\subsection{The user-defined structures and functions and their interconnections}
\begin{figure}[H]
\begin{center}
\includegraphics[width=0.8\textwidth]{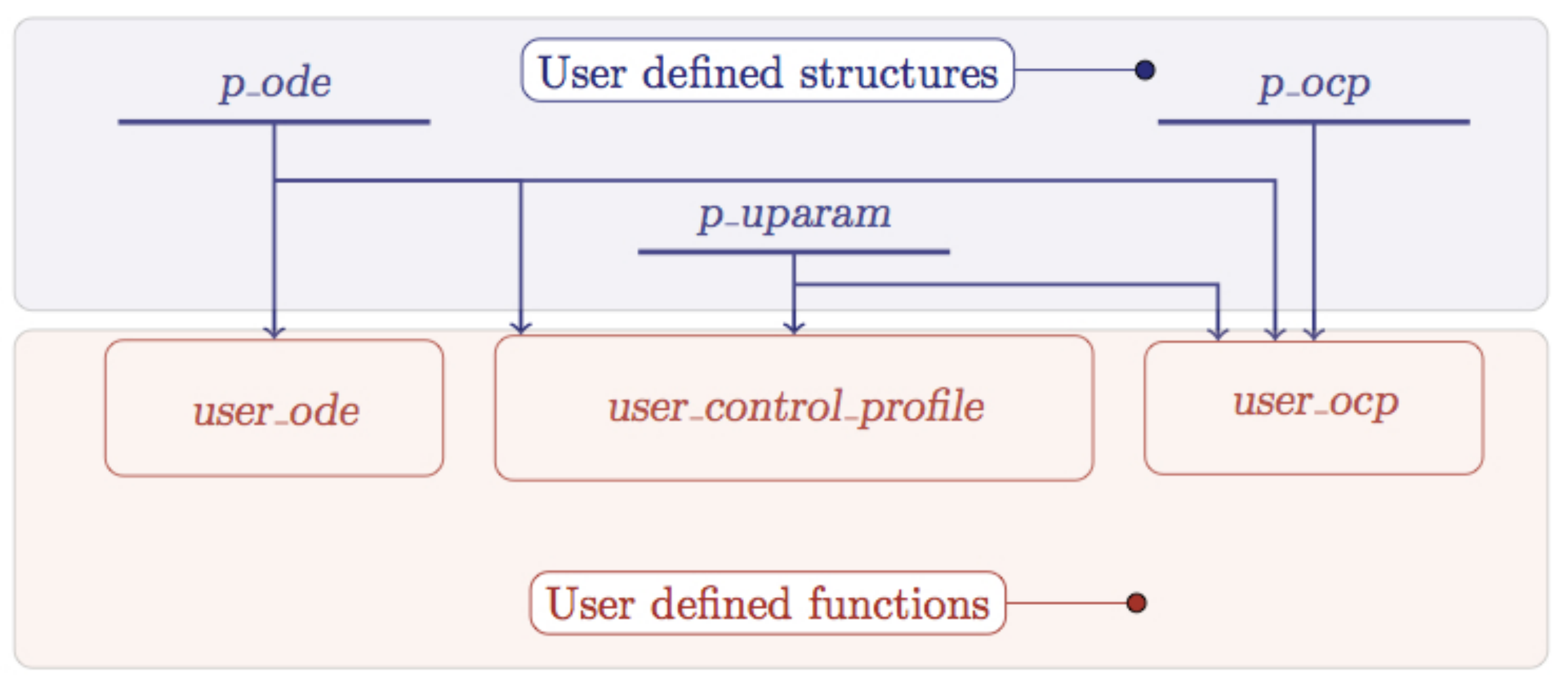} 
\end{center} 
\caption{User-defined structures and functions} \label{fig_userdefined} 
\end{figure}
Figure \ref{fig_userdefined} shows the items that have to be provided by the user. These items are divided into two categories:
\begin{enumerate}
\item {\bf The user-defined structures}: This includes the structures invoked in section \ref{basicsMPC}, namely:
\begin{itemize}
\item[$\checkmark$] \pode needed for the definition of the dynamics
\item[$\checkmark$] \puparam needed for the definition of the control profile's parametrization 
\item[$\checkmark$] \pocp needed to define the cost function and the constraints
\end{itemize}  
\item {\bf The user-defined functions}: This includes the three functions:
\begin{itemize}
\item[$\checkmark$] \userode the function that defines the dynamics [see (\ref{system})]
\item[$\checkmark$] \useruparam the function that defines the control profile's parametrization [see (\ref{uparam})]
\item[$\checkmark$] \userocp the function that defines the cost and the constraints [see (\ref{theocp})]
\end{itemize} 
\end{enumerate}  
The interaction between these structures and functions is described in Figure \ref{fig_userdefined}. \ \\ \ \\ 
Figure \ref{fig_wholepicture} shows how the user-defined items are used to build the solution. This is done in two steps: 
\begin{enumerate}
\item In the first step, a global structure called {\sl param} is created by invoking the subroutine:
\begin{center}
\cp
\end{center} 
provided by the \pac package. This creates the structure {\sl param} that gathers the user-defined structures together with other items that are needed to create the MPC feedback function. This call takes the form:
\begin{center}
[param,flag,message,teval]=\cp(\pode,\ \puparam,\ \pocp,mode) \label{previouscall} 
\end{center} 
namely, the user-defined structures are used as argument to the subroutine \cp which returns the following arguments:
\begin{figure}
\begin{center}
\includegraphics[width=0.8\textwidth]{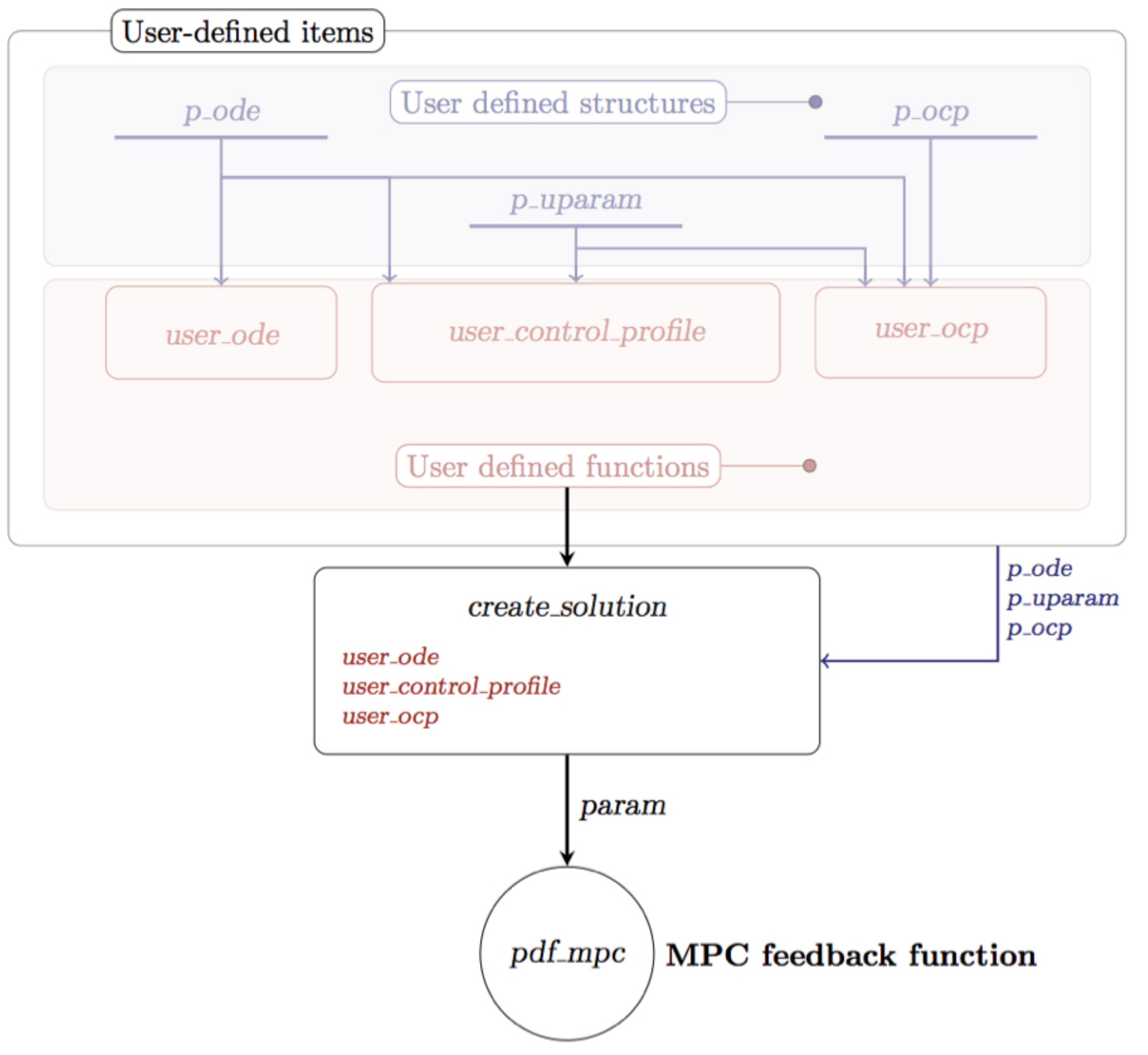} 
\end{center} 
\caption{Structure of the \pac package} \label{fig_wholepicture} 
\end{figure}
\begin{tabbing}
\hskip 15mm \= \hskip 20mm \kill 
{\sl param}  \> :the global structure invoked above,\\
flag \>:Success/failure indicator,\\
message \>:the corresponding message,\\
teval \> :the evaluation of the time needed for a single evaluation of the (cost function,constraint) pair. \\
mode \> :$\{0,1\}$-valued argument that tells {\em create\_solution} whether a re-compilation is necessary.   
\end{tabbing}
Note that the last input arguments ({\sl mode}) of the call to the {\sl create\_solution} function indicates whether a new compilation is needed or not. This enables to avoid useless re-creation of the {\sc Matlab} {\sc mex}-function {\sl pdf\_mpc} a change in the problem parameters that does not affect the compiled function.  More precisely: if {\sl mode}=1 then re-compilation is performed, otherwise, no re-compilation is performed. More detailed discussion regarding the conditions under which a recompilation is necessary is given in section \ref{whentorecompile}. 
\ \\ \ \\ 
As mentioned above, the last output (teval) is interesting in order to evaluate the number of (cost/constraint) pair evaluations to be given to the {\sl pdf\_mpc} subroutine through the Nev-field of the parameter structure {\sl param}. 
\begin{center}
{\sl param.Nev} 
\end{center} 
\item As a matter of fact, the subroutine {\sl create\_solution} builds a {\sc Matlab} {\sc Mex}-function called pdf\_mpc whose call shows the following syntaxe:
\begin{equation}
\mbox{\rm [param,u,u\_sol,t\_exec]=pdf\_mpc(x,param,subset)}  \label{callpac} 
\end{equation}   
\end{enumerate}  
where 
\begin{tabbing}
\hskip 15mm \= \hskip 20mm \kill 
$x$ \> :the current state at which the future optimal sequence of control is to be computed\\
param \> :the structure created by the call of {\sl create\_solution} (see above) \\
subset \> :an optional argument containing a subset of components of the decision variable $p$ to be optimized\\
u \> :the MPC feedback to be applied \\
u\_sol \> :the optimal sequence of control inputs that solved the constrained optimization problem. \\
t\_exec \> :the time needed to perform the param.Nev iterations. 
\end{tabbing}
Recall that  $$u=\mbox{\rm u\_sol}(1:n_u)$$ namely the first control in the optimal sequence is to be applied to the system before a second round of iterations is started at the next sampling period. The total vector u\_sol is a column vector containing the future optimal actions over the prediction horizon of length param.uparam.Np. therefore, the following command puts the sequence in a time$\times$ components form:
\begin{center}
(reshape(u\_sol,$n_u$,param.uparam.Np))'
\end{center} 
\subsection{Required mandatory fields for the user-defined structures} \label{secreqfields} 
The user-defined structures \pode, \puparam and \pocp need some required fields to be present for the construction of the param structure by the {\sl create\_solution} function. These fields are enumerated in table \ref{tabrequiredfields}. Note that any other fields that would be necessary can be used. Table \ref{tabrequiredfields} only shows those whose presence is mandatory. If the user does not provide one of these fields, an error is returned with an appropriate message. 
\begin{table}[H]
\begin{center}
\begin{tabular}{|l|l|l|} 
   \hline
    {\bf Structure} & \multicolumn{2}{l|}{\bf Required fields} \\
    \hline
    \pode & tau & sampling period \\
        & x0 & initial state \\
        & u0 & initial control \\
        & rk\_order & Runge-Kutta order \\
    \hline
    \puparam & nu & input size \\
        & Np & prediction horizon \\
        & np & dimension of $p$ \\
        & p & initial value of $p$ \\
        & pmin & lower bound on $p$ \\
        & pmax & upper bound on $p$ \\
    \hline
    \pocp & -- & -- \\
    \hline
\end{tabular}
\end{center}
\caption{The required fields for the user-defined structures  \pode, \puparam and \pocp.} \label{tabrequiredfields} 
\end{table} 
\subsection{The available useful fields in the {\sl pdf\_mpc}'s {\sl param} structure}
The structure {\sl param} created by the {\sl create\_solution} function has many useful fields that can be exploited when using the {\sl pdf\_mpc} package in designing and using real-time NMPC state feedback. These fields are enumerated and their possible use is commented. Note however that only the subset of fields that are useful are described. This is not an exhaustive description of the {\sl param}-structure fields.
\subsubsection{param.ode.rk\_order}
This is the Runke-kutta order used by the solver to compute the cost and the constraints. The possible values are $\{1,2,4\}$. Note that taking small values accelerates the computation while introducing the risk of integration error. The resulting feedback can be validated by integrating the resulting closed-loop system using a \pode structure with higher value p\_ode.rk\_order to check the implication of low precision prediction on the closed-loop behavior (see the description of the utility function {\sl one\_step} in section \ref{onestep}). 
\subsubsection{param.Nev}
This is the {\bf maximum} number of cost/constraints evaluation that can be performed by the function {\sl pdf\_mpc} that computes the MPC optimal sequence.  
\subsubsection{param.ode.u0}
This field represents the previous control applied to the system. It can be useful when increment between two successive control, namely $u(k)-u(k-1)$ is penalized and/or constrained. By default each time pdf\_mpc is invoked, this parameter is set to the control to be applied so that in the next iteration, it takes naturally the value of the previous control. However, if multi-step open-loop application of the resulting optimal sequence u\_sol computed by the pdf\_mpc function is used, {\sl param.ode.u0} has to be manually enforced to the last control being applied.  
\subsubsection{param.pmin, param.pmax}
lower and upper bounds on the decision variable $p\in \mathbb{R}^{n_p}$. The initial values of these fields are inherited from p\_uparam.pmin and p\_uparam.pmax of the structure p\_uparam used in the call of the function {\sl create\_solution}. During the closed-loop, they can be updated if necessary and the new assigned values are then used in the solver.  
\subsubsection{Trust region updating parameters}
The \pac package optimization solver uses a trust-region mechanism. A quadratic approximation of the cost function {\sc and} the constraints function is obtained over a trust-region of size, say $\alpha$. If the decision based on these approximation are successful, the size of the trust region is increased (the trust-region is expanded). Otherwise, the size is reduced (the trust-region) is contracted. \ \\ \ \\ 
The expansion rate (after a success) and the contraction rate (after a failure) are defined by two strictly positive scalars $\beta^+>1$ and $\beta^-<1$ respectively. The values of these two parameters can highly impact the rate of convergence and the precision of the results, in particular when the number of iterations is limited. Indeed, if $\beta^+$ and $\beta^-$ are taken too close to $1$, a high number of iterations would be needed but the final result would be very precise. On the other hand, taking $\beta^+\gg 1$ and/or $\beta^-\ll 1$ would lead to very rapid convergence to a loosely wide region around the optimum. The by default values are taken respectively equal to $2$ and $0.5$ which nicely work in almost all the situations I encountered. However, the user can modify these values by using the provided subroutine {\sl update\_trust\_region\_parameters} as follows:
\begin{center}
param={\sl update\_trust\_region\_parameters}(param,[$b^+$,$b^-$])
\end{center}  
This call updates the parameter of the structure {\sl param} appropriately.  
\subsubsection{param.compiled}
When the {\sl param} structure is created through the call of the function {\sl create\_solution}, the default value of {\sl param.compiled} is 1 and the compiled {\sc mex}-function of the solver is used. When   {\sl param.compiled} is set to 0, it is the interpreted {\sc matlab} version of the solver which is executed. This is much slower (by two orders of magnitude) than the compiled version. It is therefore never advised to use the value {\sl param.compiled}=0 unless the {\sc matlab}-coder is not available. The resulting solution is rarely compatible with real-time implementation of the corresponding MPC-feedback.    
\subsection{The available useful functions of the {\sl pdf\_mpc} package}
Almost all the useful functions have been already mentioned in the preceding sections. They are recalled here for easiness of references:
\subsubsection{create\_solution}
This is the function that should be called once the required user-defined structures \pode, \puparam and \pocp have been defined together with the functions \userode, \useruparam and \userocp. We have already mentioned the syntaxe of this call:
\begin{center}
[param,flag,message,teval]=\cp(\pode,\ \puparam,\ \pocp, mode) \label{previouscall} 
\end{center}
This creates the basic structure {\sl param} that is used by the solver. Moreover, depending on the  value of the {\sl mode} arguments, the compiled version of the {\sc mex}-function {\sl pdf\_mpc} is re-calculated. 
\subsubsection{pdf\_mpc}
This is the function that computes the optimal control sequence from which the first action represents the MPC feedback. Recall that this function is created by the function {\sl create\_solution}. Its call takes the following form:
\begin{equation}
\mbox{\rm [param,u,u\_sol,t\_exec]=pdf\_mpc(x,param,subset)} 
\end{equation} 
where the input and output arguments have been already discussed earlier (see the discussion following (\ref{callpac})).
\subsubsection{update\_trust\_region\_parameters}    
This is the function already mentioned that enables the user to change the expansion/contraction parameters of the trust-region mechanism used by the \pac package's optimizer. This function can be called only after the structure {\sl param} is created via the call of the function {\sl create\_solution}. The call of this function takes the following form:
\begin{center}
param={\sl update\_trust\_region\_parameters}(param,[$b^+$,$b^-$])
\end{center}    
where $b^+>1$ and $0<b^-<1$ are the expansion/contraction factors respectively. \ \\ \ \\ 
When an optional last argument is present:
\begin{center}
param={\sl update\_trust\_region\_parameters}(param,[$b^+$,$b^-$],alpha\_min)
\end{center}    
the function set also the value of the smallest size of the trust-region. Two options are then available depending on the size of the last argument:
\begin{enumerate}
\item Either alpha\_min is scalar in which case, the minimum size is used for all the components of the decision variable $p$
\item Or alpha\_min is a vector of dimension $n_p$ in which case the sizes of the trust region are set accordingly. 
\item Any other size leads to an error.  
\end{enumerate} 
The default value of alpha\_min is set to $10^{-9}$.
\subsubsection{one\_step} \label{onestep} 
This is a map that computes a one-step integration of the system's model. Its call takes the form:
\begin{center}
xplus=one\_step(x,u,p\_ode)
\end{center} 
where 
\begin{tabbing}
\hskip 15mm \= \hskip 20mm \kill 
x \> :the initial state $\in \mathbb{R}^{n_x}$\\ 
u \> :the control input $\in \mathbb{R}^{n_u}$\\ 
p\_ode \> :a structure that fits the initially created \pode strcuture.
\end{tabbing}
Note that once the {\sl param} structure is created, a field {\sl param.ode} is created by copying the user defined \pode structure. Using this field, the solver use {\sl param.ode} in the call of {\sl one\_step} to simulate the open-loop trajectory while the original \pode can be used in the simulation of the closed-loop system. This enables uncertainties to be simulated while the controller is computed on the nominal system.  
\subsubsection{simulate\_ol}
This function simulates the open-loop behavior of the system over a prediction horizon of length {\sl p\_uparam.Np} using the sampling period p\_ode.tau where p\_ode and p\_uparam are the structures given as second and third arguments while the first argument $p$ is the value of the control profile's parameter. This function can be used in the first steps of the design in order to check the model and the control parametrization before closing the loop via the optimization process. The call of {\sl simulate\_ol} takes the form: 
\begin{center}
[tt,xx,uu]=simulate\_ol(p,p\_ode,p\_uparam)
\end{center} 
where 
\begin{tabbing}
\hskip 15mm \= \hskip 20mm \kill 
tt \> :the vector of time $\in \mathbb{R}^{n_t}$\\
xx \> :the matrix of state trajectory $\in \mathbb{R}^{n_t\times n_x}$\\
uu \> :the vector of time $\in \mathbb{R}^{n_t\times n_u}$\\
\end{tabbing}
Note that the initial state used in this open-loop simulation is defined by the field p\_ode.x0
\subsubsection{compute\_R} \label{compute_R} 
\begin{figure}
\begin{center}
\includegraphics[width=0.6\textwidth]{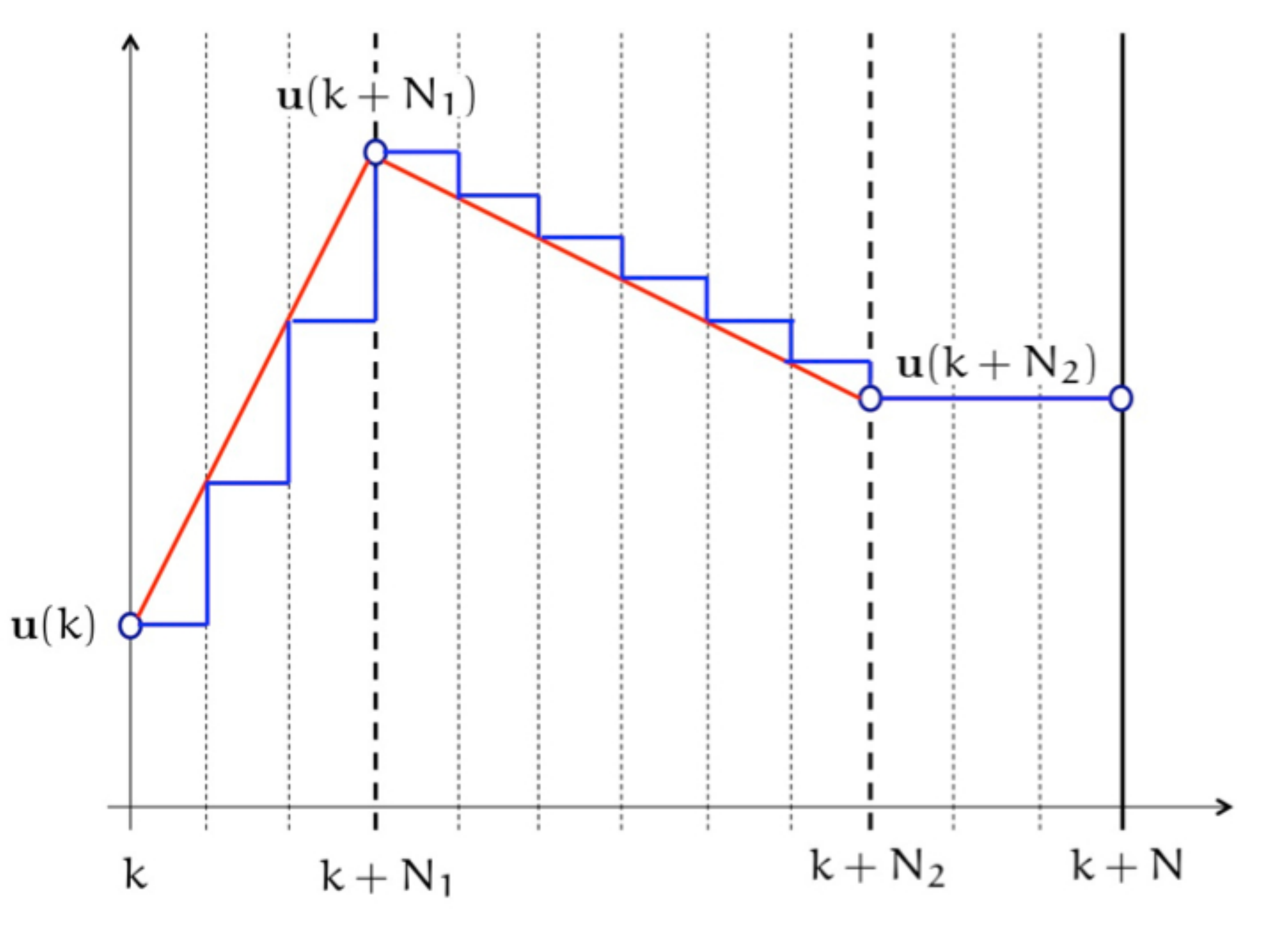} 
\end{center} 
\caption{Example of reduced parametrization associated to the useful subroutine {\sl compute\_R} provided by the \pac package. In this case, a matrix $R\in \mathbb{R}^{12\times 3}$ is returned by {\sl compute\_R} such that $\bm u=Rp$.} \label{ex_contr_param} 
\end{figure}

This utility function enables to reduce the number of degrees of freedom used in the definition of the control profile. More precisely, instead of using a piece-wise constant profile in which all the values of the control vector at all the sampling period over the prediction horizon of length $N$ are viewed as decision variables, only a subset of these sampling periods are associated to free control values, the remaining values are obtained by linear interpolation. 
\begin{center}
R=compute\_R(Ifree,N,nu)
\end{center} 
where 
\begin{tabbing}
\hskip 15mm \= \hskip 20mm \kill 
Ifree \> :the vector of indices of the sampling period with free values\\
N \> :the number of sampling periods in the prediction horizon\\
nu \> :the number of control inputs (actuator)\\
R \> :the matrix that reconstructs the control profile $\bm u$ from the decision variable $p$ according to: 
\end{tabbing}
\begin{equation}
\bm u=R\cdot p
\end{equation}
Note that if the last index is strictly lower than N=Np, namely Ifree(end)$<$Np, the tail of the control profile is supposed to be constant. Figure \ref{ex_contr_param} show a schematic of this parametrization technique. In this case, NIfree=[1,4,10] while N=12.
 
\subsubsection{initialize}
This is a utility function that is used in the creation of the matrices for the closed-loop result. It creates the $0$-matrices of appropriate size that can be later used in saving the closed-loop state, control, time, execution times. The call takes the form:
\begin{center}
[tt,xx,uu,tt\_exec,ntsim]=initialize(tsim,param);
\end{center} 
where 
\begin{tabbing}
\hskip 15mm \= \hskip 20mm \kill 
tsim \> :the duration of the closed-loop simulation\\
tt \> :the the vector of time given tsim and param.ode.tau\\
xx \> :the $0$-matrix of dimension ntsim$\times n_x$ to welcome the closed-loop state evolution\\
uu \> :the $0$-matrix of dimension ntsim$\times n_u$ to welcome the closed-loop control evolution\\
tt\_exec \> :the $0$-vector of dimension ntsim to welcome the MPC-computation times\\
ntsim \> :the number of instants in the closed-loop simulation. 
\end{tabbing}

\section{Some important issues to be understood} \label{somimpo} 
\subsection{The fields {\sl param.ode, param.uparam} and {\sl param.ocp}}
When the structure {\sl param} is created, the structures \pode, \puparam and \pocp  are copied in the fields {\sl param.ode, param.uparam} and {\sl param.ocp} respectively. These last fields are then used by the solver to simulate the system's equation, to compute the candidate profile given a candidate decision variable $p$ and to compute the corresponding cost and the constraint. During the closed-loop simulation, if the user need to change the definition of the cost function (to take an example), then it is the value of some field of the structure {\sl param.ocp} that has to be changed and not that of \pocp. In other words, as far as the solver is concerned, the structures \pode, \puparam and \pocp are only used to define the skeletons and the initial values of {\sl param.ode, param.uparam} and {\sl param.ocp}. Once this is done, the structures \pode, \puparam and \pocp no more influence the solver p\_ode can be used in the closed-loop to simulate a a different set of parameters in order to evaluate the robustness of the result to modeling errors. 
\subsection{When to recompile the solution} \label{whentorecompile} 
Once the solution {\sc mex}-function  {\sl pdf\_mpc}  is created, it can be used for different values of any subset of of its fields {\bf provided that the new values does not change the dimension of the decision variable $p$}. For instance, one can change the sampling period field {\sl param.ode.tau}, the penalties that would be defined as fields of {\sl param.ocp} such as {\sl param.ocp.Q} or {\sl param.ocp.R} and so on. However, the fields {\sl param.Np} and/or {\sl param.np} cannot be changed since such a change would imply a change in the number of decision variables. This issue could have been solved using dynamic allocation at the price of a lower performance. This option was not adopted in the current version of the \pac package. Consequently if changes are required that affect the dimension of the decision variable, then a rebuilding of the solution is necessary by using {\sl mode}=1 as a last argument of {\sl create\_solution}. The same situation occur when using {\sl compute\_R} in order to reduce the number of decision variable, namely, if the dimension of the argument {\sl Ifree} changes, a recompilation is necessary since the dimension of $p$ changes. 
\subsection{Output format for {\sl \useruparam}}
When defining the \useruparam map which is to be called according to:
\begin{equation}
\bmu:=\mbox{\rm \useruparam}(p,\mbox{\rm \pode},\mbox{\rm \puparam})  \label{uparambis} 
\end{equation} 
It is important to respect the output format returned by the user-defined function \useruparam. Namely, the output should be a matrix with as many lines as time instants (This is the fields p\_uparam.Np) and as many columns as input components (defined by the field p\_uparam.nu). More clearly, the following format should be returned:
\begin{equation}
\bm u=\begin{bmatrix}
u_1(t_1)&\dots&u_{n_u}(t_1)\cr 
u_1(t_2)&\dots&u_{n_u}(t_2)\cr 
\vdots &\vdots & \vdots \cr 
u_1(t_N)&\dots&u_{n_u}(t_N)\cr 
\end{bmatrix}  \label{formatmatrux} 
\end{equation} 
\subsection{Input arguments formats for {\sl \userocp}}
When defining the \userocp map which is called according to:

\begin{equation}
[J,g]=\mbox{\rm \userocp}(\bm x,\bmu,\mbox{\rm \pode,\puparam,\pocp}) \label{theocpbis} 
\end{equation} 
the format for the input arguments $\bm x$ and $\bm u$ must follow (\ref{formatmatrux}). 
\subsection{Nested functions}
The {\sc matlab-coder} does not support the nested functions. These are functions that are defined inside another functions. More clearly all functions needed in he definition of the used defined functions \userode, \useruparam and \userocp  must be defined in a separate .m file each. Nested functions may be possible in future versions of {\sc matlab-coder} but at the time this manual is written, this is not possible.  
\subsection{Global variables}
I never used global variables in the context of the \pac package. I strongly advise always using the user-defined structures (and not global variables) in order to feeds the parameters to the functions. 
\subsection{Hard vs soft constraints}
Hard constraints should be used only when necessary. This is because for real-system, it is difficult to known in advance whether they are always feasible or not. The \pac package concentrate on the satisfaction of hard constraints before optimizing the cost function. This means that in case where the hard constraints are unfeasible, the solver {\bf does not} necessarily give the best trade-off between constraint violation and the minimization of the cost function. It will simply tries to minimize the constraints violation indicator. This is why it is often a good action to include the constraints with appropriate penalties in the cost function. \ \\ \ \\ More precisely, assume that in the original formulation the pair $J,g$ representing the cost function and the constraints violation are delivered by the \userocp subroutine. Moreover, assume that the constraint violation indicator $g$ is defined by $g=max([g_h,g_s])$ where $g_h$ and $g_s$ are possibly hard and soft constraints contribution which are all viewed as hard constraints in the original formulation. Now a better formulation can be obtained by using the following new definitions in \userocp:
\begin{align*}
J&\leftarrow J+ \mbox{\rm p\_ocp}.penalty*max(g_s,0)^2\\
g&\leftarrow max([g_h])
\end{align*} 
where p\_ocp.penalty is the field that can be tuned so that the satisfaction of the soft constraints is acceptable over the reasonably set of realistic scenarios. Note that if no hard constraints are present, the second line can be simply replaced by:
\begin{equation*}
g\leftarrow -1
\end{equation*}
{\bf Note however that the box constraints $p\ge p_{min}$ and $p\le p_{max}$ are always handled as hard constraints.} Recall however that these constraints are not to be delivered by the \userocp subroutine.
\section{Use examples} \label{secexamples} 
The best way to learn how to use the \pac package is to go through some examples. Several examples are given to show different aspects of  the package and how flexible it is to address different situations and control parametrization. For each example, the problem is first stated in terms of model equations, constraints and control objective. The solution's script is given and then its different parts are commented. As far as the computation times are concerned, they corresponds to the package being used on a Mac PowerBook using OS X version 10.9.5 with 2.8 GHz Intel Core i7 processor and 16 Go 1600 MHz DDR3 memory ship. 
\subsection{Example 1: control of a crane}
\subsubsection{Problem statement}
Let us reconsider the example of the crane shown below 
\begin{center}
\includegraphics[width=0.35\textwidth]{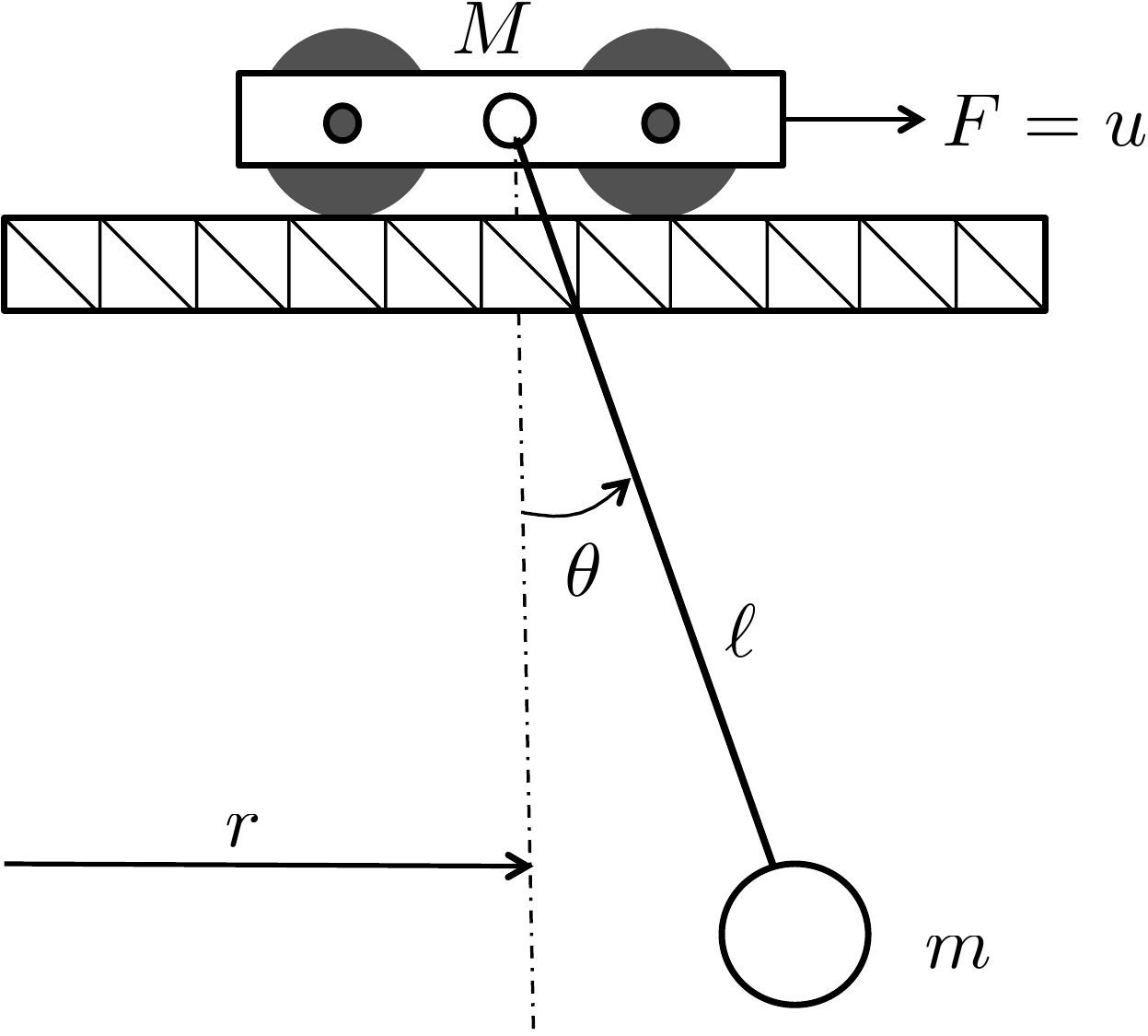}
\end{center} 
The nonlinear model of the crane can be described by the following set of ordinary differential equations:
\begin{eqnarray}
\ddot r&=&\dfrac{u+mg\cos\theta\sin\theta+m\ell \dot{\theta}^2\sin\theta}{M+m(1-\cos^2(\theta))} \label{craneee1} \\
\ddot\theta &=& \dfrac{-u\cos\theta-m\ell\dot{\theta}^2\cos\theta\sin\theta-(M-m)g\sin\theta}{(M+m/sin^2\theta)\ell} \label{craneee2} 
\end{eqnarray} 
The control objective is to steer the position of the cart $r$ to some desired position $r_d$ while meeting the following constraints:
\begin{equation}
(\theta,\dot\theta)\in [-\theta_{max},+\theta_{max}]\times [-\dot\theta_{max},+\dot\theta_{max}] \quad;\quad u\in [-u_{max},+u_{max}]
\end{equation} 
In order to achieve this task, an MPC design is used with the cost function defined by:
\begin{equation}
\sum_{k=0}^N \|x(k)-x_d\|_Q^2+\|u(k)\|_R^2+\|\Delta u(k)\|_M^2
\end{equation} 
for some weighting positive definite matrix $Q$ and two scalars $R>0$ and $M>0$. $x_d:=(r_d,0,0,0)^T$ is the steady state corresponding to te desired values $r_d$. 
\subsubsection{Solution}
Figure \ref{crane_solution} shows the complete solution to the crane problem using the \pac package. Note the script can be divided into five parts:
\begin{enumerate}
\item The definition of the \pode structure
\item The definition of the \puparam structure
\item The definition of the \pocp structure
\item The call of {\sl create\_solution}
\item The simulation of the closed-loop and the plot of the results.  
\end{enumerate}  
Note that the user-defined procedure {\sl user\_plot} appearing at the end of the script is not shown here as it is a simple successive plot instructions that show the closed-loop behavior of the variables. Figure \ref{crane_ode}, \ref{crane_control_profile} and \ref{crane_ocp} shows the scripts of the corresponding user-defines map {\sl user\_ode}, {\sl user\_control\_profile} and {\sl user\_ocp}. 
\ \\ \ \\ 
Note that there is a problem specific function called {\sl user\_sim} that is used to define the  evolution of the field {\sl param.ocp.rd} which represents the desired value of the cart's position. This script of this function is given in Figure \ref{crane_sim}.  
\begin{figure}[H]
\begin{minipage}{0.12\textwidth}
\ 
\end{minipage} 
\begin{minipage}{0.9\textwidth}
\begin{center}
\footnotesize
\color{Blue}
\begin{verbatim}
%-------------------------------------------------------------------------------
% pdf_mpc package: Example 1 - control of a crane.
%-------------------------------------------------------------------------------
% Definition of p_ode
%-------------------------------------------------------------------------------
p_ode.tau=0.5;
p_ode.rk_order=4;
p_ode.x0=[0;0;0;0];
p_ode.u0=0;
p_ode.w=[1;-0.2;-0.2];
%-------------------------------------------------------------------------------
% Definition of p_uparam
%-------------------------------------------------------------------------------
p_uparam.nu=1;
p_uparam.Np=20;
p_uparam.Ifree=[1;2;3;10];
p_uparam.R=compute_R(p_uparam.Ifree,...
    p_uparam.Np,p_uparam.nu);
p_uparam.np=size(p_uparam.R,2);
p_uparam.p=zeros(p_uparam.np,1);
p_uparam.pmin=-30*ones(size(p_uparam.R,2),1);
p_uparam.pmax=+30*ones(size(p_uparam.R,2),1);
%-------------------------------------------------------------------------------
% Definition of p_ocp
%-------------------------------------------------------------------------------
p_ocp.Q=diag([1e8;1e4;1;1]);
p_ocp.R=1e2;
p_ocp.M=1e4;
p_ocp.rd=1;
p_ocp.theta_max=0.0035;
p_ocp.thetap_max=2*pi/30;
%-------------------------------------------------------------------------------
% Create the param strcurure 
%-------------------------------------------------------------------------------
[param,flag,message,teval]=create_solution(p_ode,p_uparam,p_ocp,1);
%-------------------------------------------------------------------------------
% Closed-loop simulation
%-------------------------------------------------------------------------------
tsim=400;param.Nev=500;
[tt,xx,uu,tt_exec,ntsim]=initialize(tsim,param);
rrd=zeros(ntsim,1);
param=update_trust_region_parameters(param,[2,0.5]);
param.ode.rk_order=2;
for i=1:length(tt)-1
    disp(i/ntsim);
    param=user_sim(tt,i,param);
    [param,u,u_sol,tt_exec(i)]=pdf_mpc(xx(i,:)',param);
    uu(i,:)=u';
    rrd(i)=param.ocp.rd;
    xx(i+1,:)=one_step(xx(i,:)',u,p_ode);
end
rrd(i+1)=rrd(i);
%-------------------------------------------------------------------------------
user_plot;
%-------------------------------------------------------------------------------
\end{verbatim} 
\end{center} 
\end{minipage} 
\caption{Main script that solves the crane's control problem using the \pac package.} \label{crane_solution} 
\end{figure}

\begin{figure}[H]
\begin{minipage}{0.12\textwidth}
\ 
\end{minipage} 
\begin{minipage}{0.9\textwidth}
\footnotesize
\color{Blue}
\begin{verbatim}
%-------------------------------------------------------------------------------
% pdf_mpc package: Example 1 - Definition of the user_ode map
%-------------------------------------------------------------------------------
function xdot=user_ode(x,u,p_ode)
% x=(r,rp,theta,thetap)
w=p_ode.w;
m=200*(1+w(1));
M=1500;frot_theta=1e5*(1+w(2));frot_r=10*(1+w(3));
L=100;g=0.81;
th=x(3);thp=x(4);
c=cos(th);
s=sin(th);
xdot=zeros(4,1);
xdot(1)=x(2);
xdot(2)=(u+m*g*c*s+m*L*s*thp^2-frot_r*x(2))/(M+m*(1-c^2));
xdot(3)=x(4);
xdot(4)=(-u*c-m*L*thp^2*c*s-(M-m)*g*s-frot_theta*thp)/((M+m*s^2)*L);
return
%-------------------------------------------------------------------------------
\end{verbatim} 
\end{minipage} 
\caption{Script of the user-defined function {\sl user\_ode} for the crane example. Note the use of the field p\_ode.w that models parameter uncertainties so that when the user-defined structure \pode is used, the uncertainties are simulated while when the solver invokes this function with the structure {\sl param.ode}, the corresponding uncertainties {\sl param.ode.w} is set to $0$. Note also that one could decide also to keep the physical parameters such as $M$, $m$, etc. inside the function or introduce them as fields of \pode so that they can be changed without the need for re-compilation of {\sl pdf\_mpc} {\sc mex}-function.} \label{crane_ode} 
\end{figure}

\begin{figure}[H]
\begin{minipage}{0.12\textwidth}
\ 
\end{minipage} 
\begin{minipage}{0.9\textwidth}
\footnotesize
\color{Blue}
\begin{verbatim}
%-------------------------------------------------------------------------------
% pdf_mpc package: Example 1 - Definition of the user_control_profile map
%-------------------------------------------------------------------------------
function u_profile=user_control_profile(p,p_ode,p_uparam)
    u_profile=reshape(p_uparam.R*p,p_uparam.Np,p_uparam.nu);
end
%-------------------------------------------------------------------------------
\end{verbatim} 
\end{minipage} 
\caption{Script of the user-defined function {\sl user\_control\_profile} for the crane example. Note the use of the function {\sl compute\_R} provided by the \pac package and explained in section \ref{compute_R}.} \label{crane_control_profile} 
\end{figure}

\begin{figure}[H]
\begin{minipage}{0.12\textwidth}
\ 
\end{minipage} 
\begin{minipage}{0.9\textwidth}
\footnotesize
\color{Blue}
\begin{verbatim}
%-------------------------------------------------------------------------------
% pdf_mpc package: Example 1 - Definition of the user_ocp map
%-------------------------------------------------------------------------------
function [J,g]=user_ocp(xx,uu,p_ode,p_uparam,p_ocp)
J=0;xd=[p_ocp.rd;0;0;0];
for i=1:p_uparam.Np
    if i==1
        du=uu(1)-p_ode.u0;
    else
        du=uu(i)-uu(i-1);
    end
    e=xx(i+1,:)'-xd;
    J=J+(e'*p_ocp.Q*e+p_ocp.R*uu(i)^2+p_ocp.M*du^2);
end
h1=max(xx(:,3)-p_ocp.theta_max);
h2=max(-xx(:,3)-p_ocp.theta_max);
h3=max(xx(:,4)-p_ocp.thetap_max);
h4=max(-xx(:,4)-p_ocp.thetap_max);
g=max([h1;h2;h3;h4]);
end
%-------------------------------------------------------------------------------
\end{verbatim} 
\end{minipage} 
\caption{Script of the user-defined function {\sl user\_ocp} for the crane example. Note that the cost function is defined for a given value of the field {\sl param.ocp.rd}. This values if modified on line through the {\sl user\_sim} function shown in Figure \ref{crane_sim}.} \label{crane_ocp} 
\end{figure}

\begin{figure}[H]
\begin{minipage}{0.12\textwidth}
\ 
\end{minipage} 
\begin{minipage}{0.9\textwidth}
\footnotesize
\color{Blue}
\begin{verbatim}
%-------------------------------------------------------------------------------
% pdf_mpc package: Example 1 - time varying param.ocp.rd
%-------------------------------------------------------------------------------
function param=user_sim(tt,i,param_past)
param=param_past;
cond1=(tt<=tt(end)/3);
cond2=(tt>tt(end)/3).*(tt<=tt(end)*2/3);
cond3=(tt>tt(end)*2/3).*(tt<=tt(end));
param.ocp.rd=cond1(i)-3*cond2(i)+3*cond3(i);
end
%-------------------------------------------------------------------------------
\end{verbatim} 
\end{minipage} 
\caption{Script of the user-defined function {\sl user\_ocp} for the crane example. This function is called at each sampling period in order to update the value of the field {\sl param.ocp.rd} that is used in the definition of the cost function inside the user-defined function {\sl user\_ocp} shown in Figure \ref{crane_ocp}.} \label{crane_sim} 
\end{figure}

\subsubsection{Results \& discussion}

\begin{figure}[H]
\begin{center}
\includegraphics[width=0.8\textwidth]{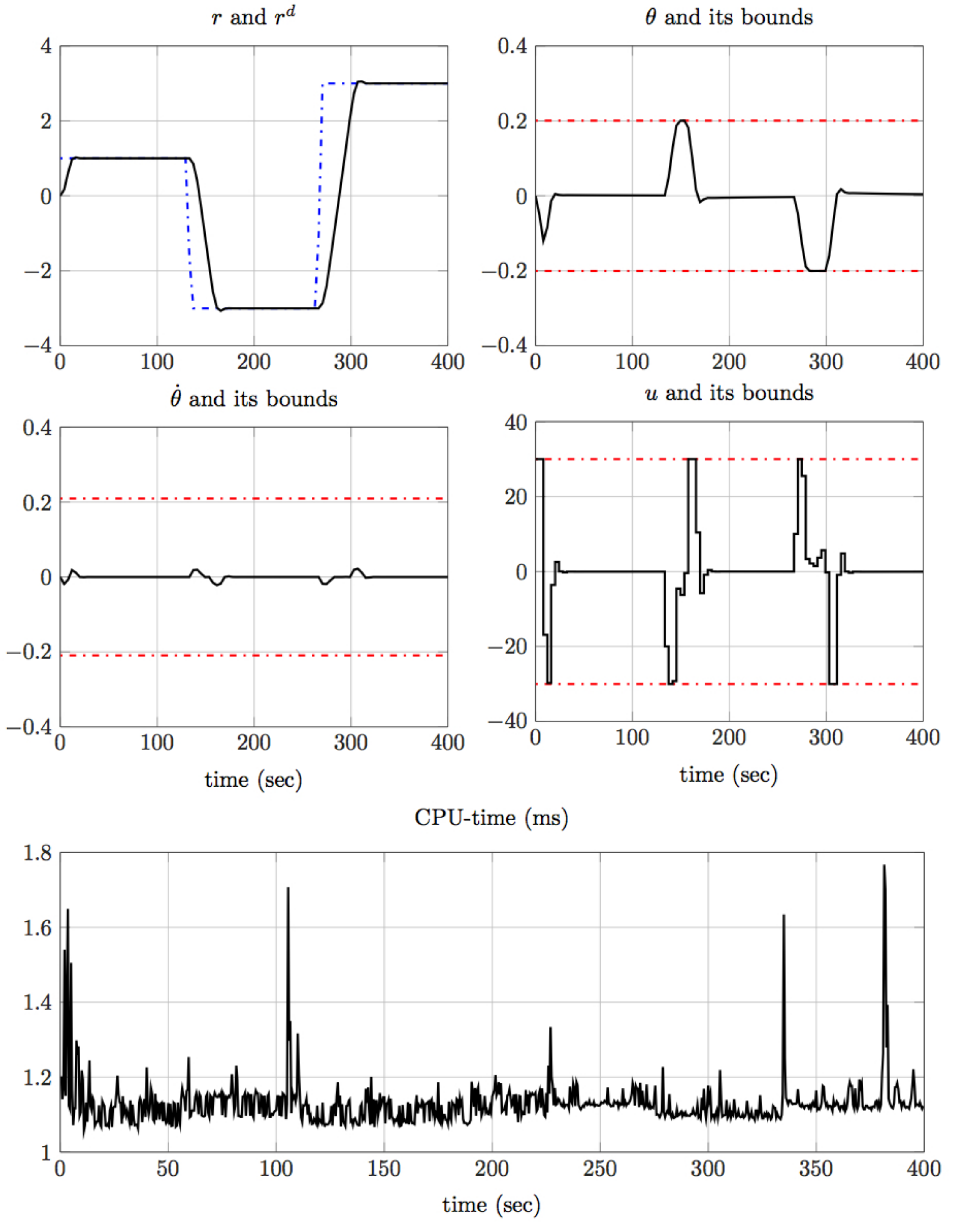} 
\end{center} 
\caption{Closed-loop simulation using the script depicted in Figure \ref{crane_solution}. Note that since the control parametrization uses {\sl param.Ifree}=\{1,2,3,10\}, the decision variable is of dimension $4$ while the prediction horizon if of dimension $20$. Figure \ref{crane_r2} shows almost indistinguishable performance with drastically shorter computation time by using the optional last argument of the {\sl pdf\_mpc} function with subset=\{1\} and by reducing the maximum number of iteration {\sl param.Nev}.} \label{crane_r1} 
\end{figure}

\begin{figure}[H]
\begin{center}
\includegraphics[width=0.8\textwidth]{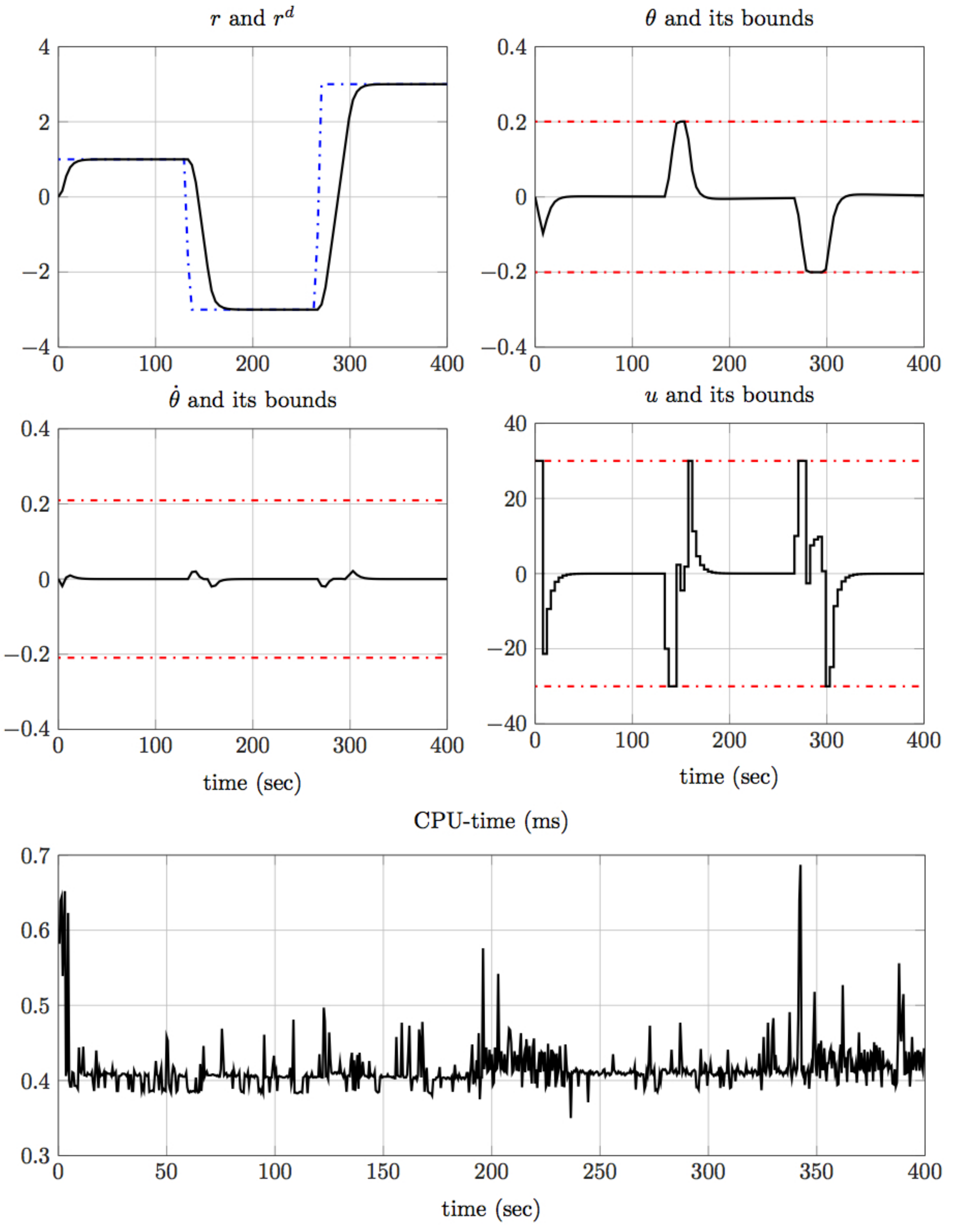} 
\end{center} 
\caption{Closed-loop simulation using the script depicted in Figure \ref{crane_solution} with the closed-loop simulation part modified as shown in Figure \ref{crane_script_modified}. Note that since the control parametrization uses {\sl param.Ifree}=\{1,2,3,10\} and the subset=[1] optional argument of pdf\_mpc is used, the decision variable is of dimension $1$. This enables to get good results while decreasing the maximum number of evaluation to {\sl param.Nev}=200.} \label{crane_r2} 
\end{figure}

\begin{figure}[H]
\begin{minipage}{0.12\textwidth}
\ 
\end{minipage} 
\begin{minipage}{0.9\textwidth}
\footnotesize
\color{Blue}
\begin{verbatim}
...
param.ode.rk_order=2;
subset=[1];
for i=1:length(tt)-1
    disp(i/ntsim);
    param=user_sim(tt,i,param);
    [param,u,u_sol,tt_exec(i)]=pdf_mpc(xx(i,:)',param,subset);
    uu(i,:)=u';
    rrd(i)=param.ocp.rd;
    xx(i+1,:)=one_step(xx(i,:)',u,p_ode);
end
...
\end{verbatim} 
\end{minipage} 
\caption{Modification of the script of Figure \ref{crane_solution} as indicated in the caption of Figure \ref{crane_r2} in order to drastically reduce the complexity and hence the computation time of the MPC feedback function {\sl pdf\_mpc}.} \label{crane_script_modified} 
\end{figure}
\newpage 
\subsection{Example 2: Combined therapy of cancer}
\subsubsection{Problem statement}
\begin{figure}[H]
\begin{center}
\includegraphics[width=0.48\textwidth]{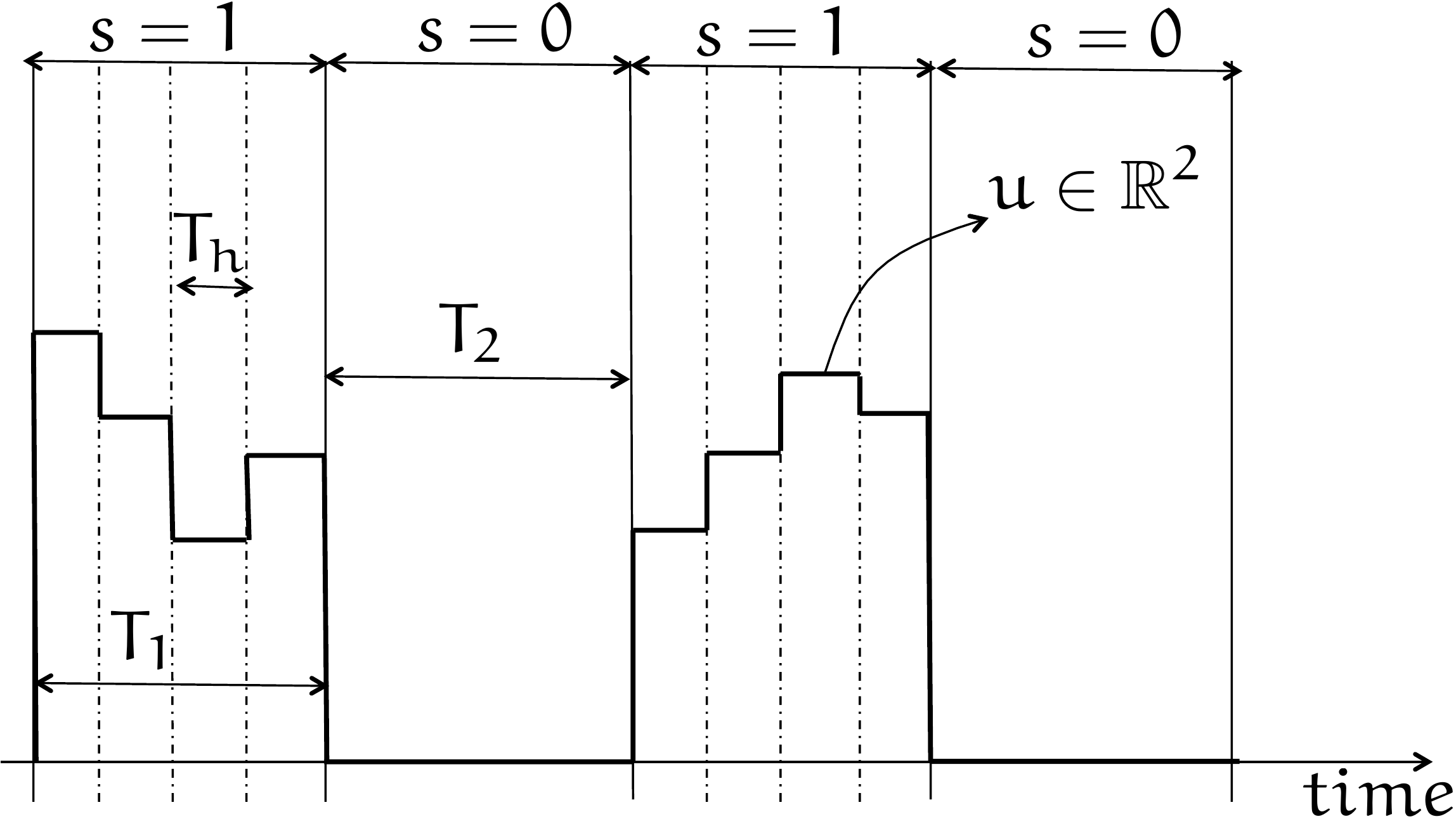} 
\end{center} 
\caption{The time structure of the drug delivery. Successive period of treatment and rest have to be observed.} \label{figcancerparam} 
\end{figure}

et us consider the following nonlinear model of cancer therapy using combined action of chemotherapy and immunotherapy:
\begin{eqnarray}
\dot x_1&=&g\dfrac{x_4}{h+x_4}x_1-rx_1-px_1x_4-k_1x_1x_3+s_1u_1 \label{cancer1}\\
\dot x_2&=&-\delta x_2-k_2x_3x_2+s_2 \label{cancer2}\\
\dot x_3&=&-\gamma x_3+u_2\label{cancer3}\\
\dot x_4&=&ax_4(1-bx_4)-c_1x_1x_4-k_3x_3x_4 \label{cancer4} 
\end{eqnarray} 
where 
\begin{tabbing}
\hskip 15mm \= \hskip 20mm \kill 
$x_1$ \> :the effector-immune cell population\\
$x_2$ \> :the circulating lymphocytes population\\
$x_3$ \> :the chemotherapy drug concentration \\
$x_4$ \> :the tumor cell population \\
$u_1$ \> :the rate of injection of the external effector-immune cells\\
$u_2$ \> :the rate of introduction of the chemotherapy drug.
\end{tabbing}
The signification of the different terms involved in (\ref{cancer1})-(\ref{cancer4}) can be found in \cite{alamir2013pragmatic}. \ \\ \ \\ 
The aim of the drug delivery is to reduce the total number of tumor cells population $x_4$ while keeping the circulating lymphocytes population $x_2$ above some lower bound $\rho$. Indeed, this population can be viewed as an indicator of health.\ \\ \ \\ 
This objective can be expressed by defining the cost function over a prediction horizon of length $N_p$:
\begin{equation}
J=x_4(N_p)
\end{equation} 
together with the associated constraint $g\le 0$ where:
\begin{equation}
g=\rho-\left[\max_{i\in \{1,\dots,N_p\}} x_2(i)\right]
\end{equation} 
The specificity of this problem lies in the time-structure constrained control profile. More precisely, the  drug delivery protocol must fit the structure depicted in Figure \ref{figcancerparam}:\ \\
\ \\
During treatment periods ($s=1$) piecewise constant control can be applied with a sampling period $T_h$. This period duration is $T_1=N_1T_h$. After, a rest period of duration $T_2=N_2T_h$ is observed during which no treatment is applied. This time structure is repeated. \ \\ \ \\ 
In terms of the control parametrization, this implies that the time structure of the control profile is time varying and an indicator needs to be introduced that defines where we are within the total period of length $T_1+T_2$. This role is played by the field {\sl p\_uparam.index} in the forthcoming programs. This is a typical non standard formulation that is not easy to handle with common piece-wise constant control parametrization\footnote{In fact, it can be handled at the price of additional equality constraints whose structure is time varying.}.
\subsubsection{Solution}
Figure \ref{cancer_solution} shows the main script that solve the problem stated in the preceding section using the {\sl pdf\_mpc} package. Note the same structure as the one used in the preceding example, namely, the definition of the structures \pode , \puparam and \pocp followed by the creation of the {\sc mex}-function {\sl pdf\_mpc} together with the main parameter structure {\sl param}.\ \\ \ \\ 
Note that the time unit here is the day so that a sampling time p\_ode.tau=0.25 means that the drug dosage is recomputed each 6 hours. Note also that the fields {\sl p\_uparam.N1} and {\sl p\_uaram.N2} represents the number of sampling period in $T_1$ and $T_2$ and that the prediction horizon used in the MPC is equal to 
\begin{center}
{\sl p\_uparam.Np=p\_uparam.N1+p\_uparam.N2}.    
\end{center} 
Note however that the number of degrees of freedoms is equal to {\sl p\_uparam.np=2(p\_uparam.N1}) because during the rest period, the control is forced to $0$. \ \\ \ \\ 
the user-defined functions \userode, \useruparam and \userocp are shown in Figure \ref{cancer_ode}, \ref{cancer_uparam} and \ref{cancer_ocp}.
\begin{figure}[H]
\begin{minipage}{0.12\textwidth}
\ 
\end{minipage} 
\begin{minipage}{0.9\textwidth}
\footnotesize
\color{Blue}
\begin{verbatim}
function xdot=user_ode(x,u,p_ode)
xdot=zeros(4,1);
xdot(1)=p_ode.g*x(4)*x(1)/(p_ode.h+x(4))-p_ode.r*x(1)-p_ode.p*x(1)*x(4)-...
        p_ode.k1*x(1)*x(3)+p_ode.s1*u(1);
xdot(2)=-p_ode.delta*x(2)-p_ode.k2*x(3)*x(2)+p_ode.s2;
xdot(3)=-p_ode.gam*x(3)+u(2);
xdot(4)=p_ode.a*x(4)*(1-p_ode.b*x(4))-p_ode.c1*x(1)*x(4)-p_ode.k3*x(3)*x(4);
end
\end{verbatim} 
\end{minipage} 
\caption{Script of the user-defined function {\sl user\_ode} for the cancer combined therapy example.} \label{cancer_ode} 
\end{figure}

\begin{figure}[H]
\begin{minipage}{0.12\textwidth}
\ 
\end{minipage} 
\begin{minipage}{0.9\textwidth}
\footnotesize
\color{Blue}
\begin{verbatim}
function u_profile = user_control_profile(p,p_ode,p_uparam)
N1=p_uparam.N1;N=p_uparam.Np;
i=p_uparam.index;
u_profile=zeros(2,N,1);
p=reshape(p,2,N1);
if i==1
    u_profile(:,1:N1)=reshape(p,2,N1);
elseif (i<=N1)
    u_profile(:,1:N1-i+1)=p(:,1:N1-i+1);
    u_profile(:,N-i+2:N)=p(:,N1-i+2:end);
else
    u_profile(:,N-i+2:N-i+N1+1)=p;
end
u_profile=u_profile';
end
\end{verbatim} 
\end{minipage} 
\caption{Script of the user-defined function {\sl user\_control\_profile} for the cancer combined therapy example.} \label{cancer_uparam} 
\end{figure}

\begin{figure}[H]
\begin{minipage}{0.12\textwidth}
\ 
\end{minipage} 
\begin{minipage}{0.9\textwidth}
\footnotesize
\color{Blue}
\begin{verbatim}
function [J,g]=user_ocp(xx,uu,p_ode,p_uparam,p_ocp)
J=xx(end,4)*(xx(end,4)>p_ocp.threshold)+(xx(end,4)<=p_ocp.threshold)*sum(sum(uu));
g=max(p_ocp.rho-xx(:,2));
end
\end{verbatim} 
\end{minipage} 
\caption{Script of the user-defined function {\sl user\_uparam} for the cancer combined therapy example.} \label{cancer_ocp} 
\end{figure}

\begin{figure}
\begin{minipage}{0.12\textwidth}
\ 
\end{minipage} 
\begin{minipage}{0.9\textwidth}
\begin{center}
\footnotesize
\color{Blue}
\begin{verbatim}
%-------------------------------------------------------------------------------
% pdf_mpc package: Example 2 - combined cancer therapy.
%-------------------------------------------------------------------------------
% Definition of p_ode
%-------------------------------------------------------------------------------
p_ode.tau=0.25;
p_ode.rk_order=4;
p_ode.x0=[5e8;1e9;0;1e9];
p_ode.u0=[0;0];

p_ode.a=25e-2;p_ode.b=1.02e-14;p_ode.c1=4.41e-10;
p_ode.f=4.12e-2;p_ode.g=1.5e-2;p_ode.r=4e-2;
p_ode.h=2.02e1;p_ode.k2=6e-1;p_ode.k3=6e-1;
p_ode.k1=8e-1;p_ode.p=2e-11;p_ode.s1=1.2e7;
p_ode.s2=7.5e6;p_ode.delta=1.2e-2;p_ode.gam=9e-1;
%-------------------------------------------------------------------------------
% Definition of p_uparam
%-------------------------------------------------------------------------------
p_uparam.nu=2;
p_uparam.N1=fix(5/p_ode.tau);
p_uparam.N2=fix(4/p_ode.tau);
p_uparam.umax=[10;1];
p_uparam.Np=p_uparam.N1+p_uparam.N2;
p_uparam.index=1;
p_uparam.np=2*p_uparam.N1;
p_uparam.p=zeros(p_uparam.np,1);
p_uparam.pmin=zeros(p_uparam.np,1);
p_uparam.pmax=reshape(p_uparam.umax*ones(1,p_uparam.N1),p_uparam.np,1);
%-------------------------------------------------------------------------------
% Definition of p_ocp
%-------------------------------------------------------------------------------
p_ocp.rho=5e7;
p_ocp.threshold=1e-40;
%-------------------------------------------------------------------------------
% Create the param strcurure 
%-------------------------------------------------------------------------------
[param,flag,message,teval]=create_solution(p_ode,p_uparam,p_ocp,0);
%-------------------------------------------------------------------------------
% Closed-loop simulation
%-------------------------------------------------------------------------------
%%
tsim=20*p_uparam.Np*p_ode.tau;param.Nev=2000;
[tt,xx,uu,tt_exec,ntsim]=initialize(tsim,param);
param=update_trust_region_parameters(param,[2,0.5]);
param.ode.rk_order=4;
for i=1:length(tt)-1
    disp(i/ntsim);
    [param,u,u_sol,tt_exec(i)]=pdf_mpc(xx(i,:)',param);
    if param.uparam.index==param.uparam.Np
        param.uparam.index=1;
    else
        param.uparam.index=param.uparam.index+1;
    end
    uu(i,:)=u';
    xx(i+1,:)=one_step(xx(i,:)',u,p_ode);
end
%-------------------------------------------------------------------------------
user_plot;
\end{verbatim} 
\end{center} 
\end{minipage} 
\caption{Main script that solves the combined cancer-therapy using the \pac package.} \label{cancer_solution} 
\end{figure}
\ \\ \ \\ 
Note that in the definition of the cost function used in the user-defined function \userocp, the field {\sl p\_ocp.threshold} is used to switch between the original objective (reducing the tumor size) to the one consisting in minimizing the drug delivered. If such definition is note used, the drug will be still delivered even if the tumor physiologically disappeared. This is because, numerically, a tumor size of $10^{-31}$ is still lower than a tumor size of $10^{-30}$. The field {\sl p\_ocp.threshold} is used here to switch off the control in that case. 
\subsubsection{Results \& discussion}
\begin{figure}[H]
\begin{center}
\includegraphics[width=0.8\textwidth]{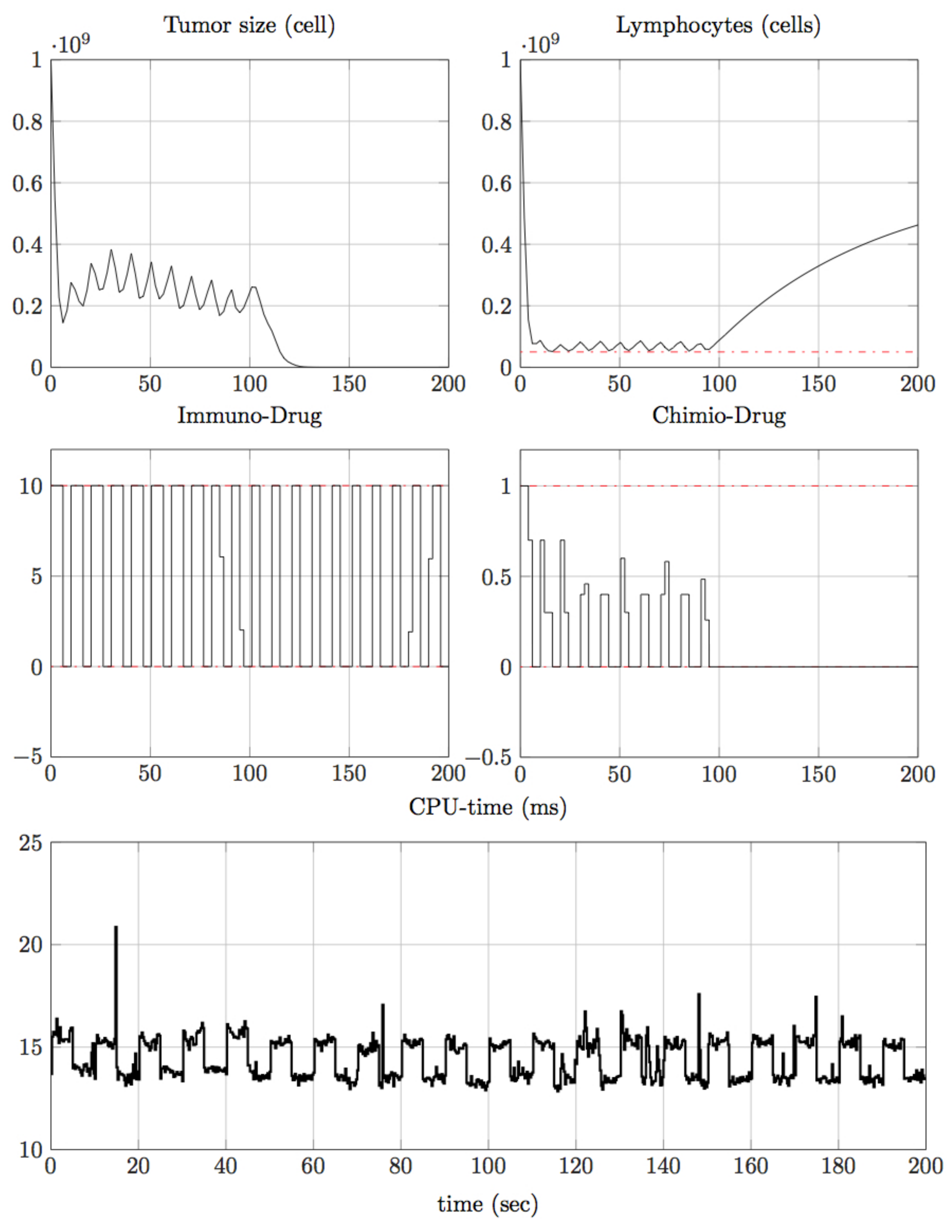} 
\end{center} 
\caption{Example 2: closed-loop evolution with $N_1=5$, $N_2=5$ and a maximum immunotherapy  drug's injection $u_1^{max}=10$} \label{cancer_r1} 
\end{figure}

\begin{figure}[H]
\begin{center}
\includegraphics[width=0.8\textwidth]{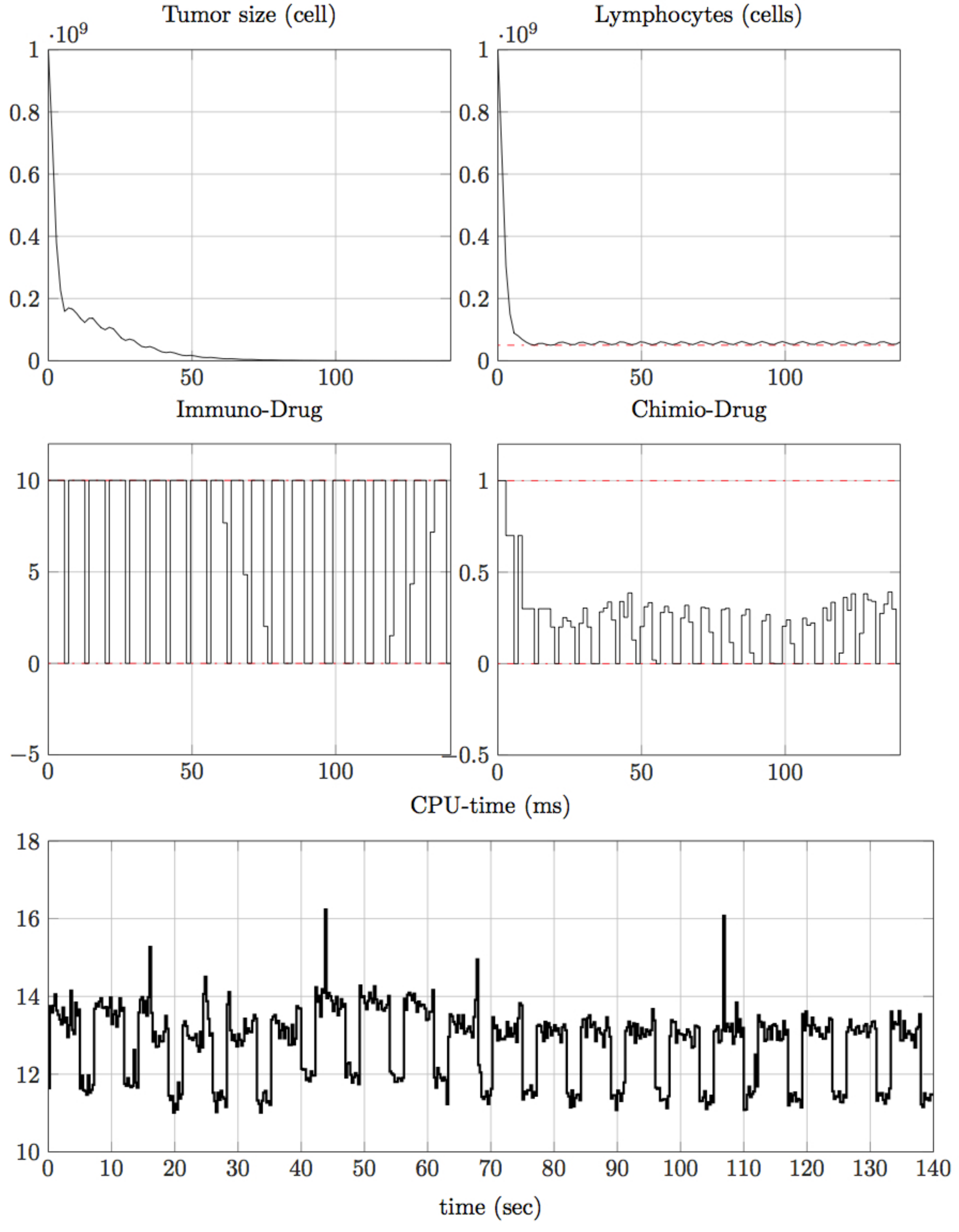} 
\end{center} 
\caption{Example 2: closed-loop evolution with $N_1=5$, $N_2=2$ and a maximum immuno drug's injection $u_1^{max}=10$} \label{cancer_r3} 
\end{figure}

\begin{figure}[H]
\begin{center}
\includegraphics[width=0.8\textwidth]{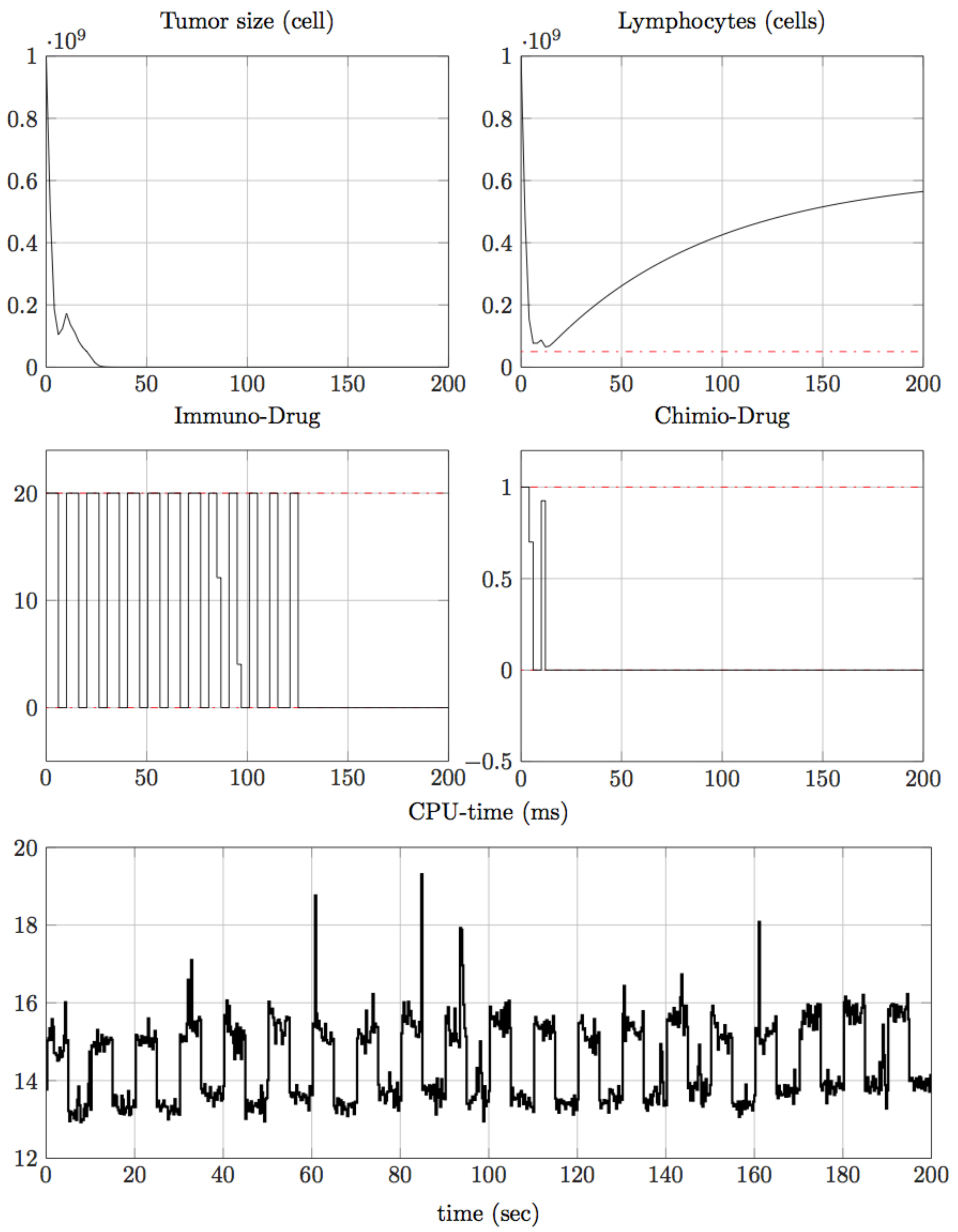} 
\end{center} 
\caption{Example 2: closed-loop evolution with $N_1=N_2=5$ and maximum immuno drug's injection of $u_1^{max}=20$} \label{cancer_r2} 
\end{figure}

Three scenarios are shown in Figures \ref{cancer_r1}, \ref{cancer_r3} and \ref{cancer_r2}. 
\begin{itemize}
\item[$\checkmark$] In Figure \ref{cancer_r1}, a scenario with $N_1=N_2=5$ days is used together with a maximum allowed injection rate of immunotherapy drug equal to 10. It can be seen that in this case, after four months, the tumor size vanishes. 
\item[$\checkmark$] In Figure \ref{cancer_r3}, only the rest period $N_2=3$ is reduced. This induces a faster decrease of the tumor size (within roughly two months).
\item[$\checkmark$] In Figure \ref{cancer_r2}, the configuration $N_1=N_2=5$ is again used but the maximum injection of immuno-therapy is doubled to $20$. In this case, the tumor vanishes in one month and the drug delivery stops as the tumor size goes below the threshold defined by the field {\sl p\_ocp.threshold} which induces the end of the treatment.  
\end{itemize} 
\newpage 

\newpage 
\appendix
\section{Downloading \& installation} \label{asecdownload} 
The \pac package can be downloaded via the {\bf software/pdf\_mpc} section of the website:
\begin{center}
http://www.mazenalamir.fr
\end{center} 
where the downloading instructions and form are provided. \ \\ \ \\ 
The \pac package is contained in a zip-file that have to be decompressed in the working folder. This package contains all the necessary files in a {\sc matlab} .p format\footnote{in order to avoid any erroneous changes by the user during the development phase and errors corrections But also to get feedback about possible errors and/or improving suggestions.} except for the user-defined files \userode, \useruparam and \userocp which are given in a skeleton .m format for the user to fill them with his/her appropriate code.  
\section{Terms of use} \label{asecterms} 
This free-software is provided with {\sc no warranty}. All consequences of its use on real-life systems is the responsibility of the user. Under these terms, the user is free to use the executable MPC solver for academic purposes provided that the citations below are included in any publication and/or public presentation of the results obtained with the \pac package: 

\begin{verbatim}
@book{alamir2013pragmatic,
  title={A Pragmatic Story of Model Predictive Control: 
  Self-Contained Algorithms and Case-Studies},
  author={Alamir, M.},
  year={2013},
  publisher={CreateSpace Independent Publishing Platform}
}
\end{verbatim} 

\begin{verbatim}
@misc{pdf_mpc,
  author = {Alamir, M.},
  title = {{A Free-Matlab-Coder package for Real-Time Nonlinear Model Predictive. 
  ar{X}iv:1703.08255},
  year = {2017}, 
}
\end{verbatim} 

\bibliography{bibmanual.bib}

\begin{thebibliography}{10}

\bibitem{Mayne2000}
D.~Q. Mayne, J.B. Rawlings, C.~V. Rao, and P.~O.~M. Scokaert.
\newblock Constrained model predictive control: Stability and optimality.
\newblock {\em Automatica}, 36:789--814, 2000.

\bibitem{Houska2011a}
B.~Houska, H.J. Ferreau, and M.~Diehl.
\newblock {ACADO} {T}oolkit -- {A}n {O}pen {S}ource {F}ramework for {A}utomatic
  {C}ontrol and {D}ynamic {O}ptimization.
\newblock {\em Optimal Control Applications and Methods}, 32(3):298--312, 2011.

\bibitem{alamir:hal-00113043}
M.~Alamir.
\newblock {\em {Stabilization of Nonlinear Systems Using Receding-Horizon
  Control Schemes: A Parametrized Approach for Fast Systems}}.
\newblock {Springer}, 2006.
\newblock Lecture Notes in Control and Identification Sciences, number 339.
  ISBN : 1-84628-470-8.

\bibitem{alamir2013pragmatic}
M.~Alamir.
\newblock {\em A Pragmatic Story of Model Predictive Control: Self-Contained
  Algorithms and Case-Studies}.
\newblock CreateSpace Independent Publishing Platform, 2013.

\bibitem{alamir2016}
M.~Alamir.
\newblock A state-dependent updating period for certified real-time model
  predictive control.
\newblock {\em IEEE Transactions on Automatic Control}, 2016.

\bibitem{diehl2005real}
M.~Diehl, H.~G. Bock, and J.~P. Schl{\"o}der.
\newblock A real-time iteration scheme for nonlinear optimization in optimal
  feedback control.
\newblock {\em SIAM Journal on control and optimization}, 43(5):1714--1736,
  2005.

\bibitem{Zavala200986}
V.~M. Zavala and L.~T. Biegler.
\newblock The advanced-step \{NMPC\} controller: Optimality, stability and
  robustness.
\newblock {\em Automatica}, 45(1):86 -- 93, 2009.

\bibitem{ALAMIR201565}
M.~Alamir.
\newblock From certification of algorithms to certified {MPC}: The missing
  links.
\newblock {\em IFAC-PapersOnLine}, 48(23):65 -- 72, 2015.

\bibitem{Alamir_ECC2013}
M.~Alamir.
\newblock Monitoring control updating period in fast gradient based {NMPC}.
\newblock In {\em European Control Conference (ECC), 2013 European}, pages
  3621--3626, July 2013.

\bibitem{pdf_mpc2017}
M.~Alamir.
\newblock {The {\sc pdf-MPC} Package: A Free-Matlab-coder package for real-time
  nonlinear model predictive control}.
\newblock \url{http://www.mazenalamir.fr/}, 2017.

\end{thebibliography}
\bibliographystyle{unsrt}
\end{document}